# Thermomechanical investigation of silicon wafer dynamics within the melting regime driven by picosecond laser pulses for surface structuring

Helen Papadaki [1,2], Inam Mirza [3], Nadezhda M. Bulgakova [3], Evaggelos Kaselouris [1,2], Vasilis Dimitriou [1,2]

[1] Physical Acoustics and Optoacoustics Laboratory, Department of Music Technology and Acoustics, Hellenic Mediterranean University, 74133 Rethymnon, Greece
[2] Institute of Plasma Physics and Lasers-IPPL, University Research and Innovation Centre, Hellenic Mediterranean University, 74150 Rethymnon, Greece
[3] FZU—Institute of Physics of the Czech Academy of Sciences, 182 00 Prague, Czech Republic

**Abstract**

Laser-induced periodic surface structures (LIPSS) on silicon, generated by ultrashort pulsed lasers, provide an efficient means to tailor surface functionality. This work presents a multiphysics finite element study on the thermomechanical dynamics of silicon wafers irradiated by picosecond laser pulses, focusing on the melting regime where thermomechanical and hydrodynamic effects dominate. To illustrate the sequential nature of laser scanning, single-pulse irradiation models are developed as thermomechanical analogues of double-pulse interactions. By positioning the laser focus near reflective boundaries and corners of the target, these models reproduce the stress-wave interference that would occur between successive pulses in scanning. The results show that periodic surface structures originate from mechanical standing wave interference within the molten layer, forming ripples with near-wavelength periodicity. The penetration depth (PD) is identified as a key factor controlling the duration and stability of these ripples: shallow PDs (75–150 nm) yield distinct, persistent patterns, while deeper PDs (~2.5 μm) lead to extended melting and hydrodynamic smoothing. Simulations of sequential double-pulse irradiation confirm that residual stresses and strains from the first pulse amplify deformation during the second, enhancing ripple amplitude and uniformity. Controlled excitation of mechanical standing waves—governed by PD, boundary geometry, and pulse sequencing—thus represents a fundamental mechanism for deterministic LIPSS formation on silicon.

## 1. Introduction

Laser-induced periodic surface structures (LIPSS) are quasi-periodic micro- or nanostructures that form on a material's surface upon irradiation with laser light, particularly from ultrashort pulsed lasers [1-4]. They consist of repetitive ripple-like patterns with a periodicity typically close to the laser light's wavelength. LIPSS can be generated on a wide variety of materials, including metals, semiconductors, and polymers [3,6-10]. The formation of ripples on the material's surface alters the properties of the affected area, including its roughness, wetting behavior, and optical absorption coefficient. The induction of LIPSS is explained by combining two fundamental approximations, the interference of surface electromagnetic waves (SEWs) and the matter reorganization. According to the first approach, the incoming laser pulse, the SEWs generated within the irradiated spot, and the SEWs scattered from the crater edge formed by previous pulses or due to surface roughness interfere, developing a local energy modulation that is imprinted onto the material via absorption [11-13]. The second approach attributes LIPSS formation significantly to hydrodynamic flow during the material's melting phase and matter reorganization during re-solidification [11,12,14].

The formation of LIPSS driven by ultrashort pulsed laser irradiation presents a promising and efficient method for tailoring surface properties. While LIPSS formation has been extensively studied with femtosecond (fs) pulsed lasers [15-19], the role of picosecond (ps) pulsed lasers in this process lacks systematic investigations [20,21]. Laser pulses of ps duration offer a unique thermal regime since they are long enough to allow for electron-lattice coupling and melting and yet short enough to minimize severe thermal damage, allowing a fascinating interplay between photophysical and photothermal processes that dictates the final surface morphology.

Numerical simulations represent a powerful methodology for investigating laser-matter interactions. By employing appropriate laser parameters and material properties, these simulations can accurately predict resultant structural and property modifications. Extensive research has utilized numerical models to elucidate the behavior of solids under laser irradiation, with a significant focus on silicon (Si) wafers subjected to ultrashort laser pulses [22-24]. In our previous works [25-27], we developed and validated three-dimensional (3D) multiphysics finite element method (FEM) models, integrating coupled thermo-structural physics, to describe the response of thin metal films and Si targets exposed to nanosecond (ns) laser pulse irradiation. In [28], we observed that regular LIPSS can be produced on Si surfaces upon ps laser scanning at rather small overlapping between irradiation spots via experiments which are representatively depicted in Figure 1, and corresponding simulations validated the results. Specifically, scattering from the laser-modified regions considering the hydrodynamic effects was successfully simulated, linking the LIPSS formation with the modification fingerprints left by each successive laser pulse. Furthermore, the experimental campaign performed on laser irradiation of Si targets using infrared (1064 nm) ps pulsed laser with a Gaussian spatial beam profile, effective focal spot diameter ($2w_o$) of laser on the target surface of ~25 μm, and peak fluence on target surface of ~2 J/cm$^2$ led to findings of equivalent importance. The post-processing of the irradiated Si samples determined that LIPSS within adjacent irradiation spots start to appear at about 19-20 mm of laser focal spot centre distances, and the amplitude of the ripples starts to increase and become more prominent with reduced inter spot distance [28]. It should be noted that similar experiments performed for molybdenum and titanium samples did not demonstrate such a replication effect. This can relate to the difference in the quality factors Q of these materials [29,30] among which Si exhibits an extraordinary Q factor at enhanced temperatures [31] thus, low-dissipating acoustic waves in Si, via their interference, can contribute to the LIPSS formation.

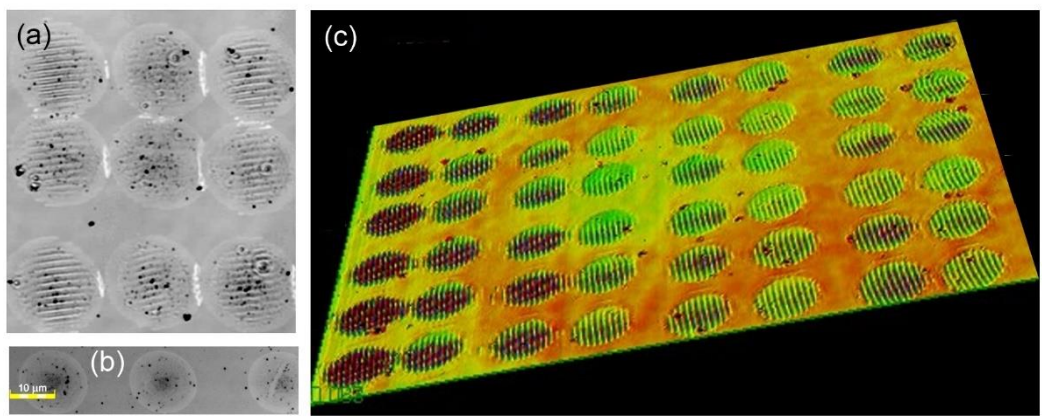

**Figure 1.** (a) Optical image of an array of irradiation spots with 1/e$^2$ diameter of 25 μm and the distance between spot centers of ~16 μm. Note that the precision of the galvo scanner (about ±2.5 μm on the sample surface) did not allow for a perfect spacing between the spots [28]. (b) Typical image of the irradiation spots when the distance between spot centers was increased to 25 μm with absence of LIPSS signs. (c) A slanted view of the LIPSS array obtained under the conditions of (a). All three images have been obtained at laser fluence of 1.04 J/cm$^2$.

The 3D FEM model is discretized by a highly refined uniform mesh to accurately capture the dynamic response of the solid Si target, enabling the detailed representation of surface modifications during and after the laser-target interaction [27,28]. To describe the behavior of the Si target, a

Johnson–Cook material model is adopted, accounting for elastoplastic effects and phase change induced by the high temperatures generated during irradiation. The hydrodynamic and bulk response of the material is modeled using a Grüneisen equation of state. The low values of density, thermal expansion coefficient, and absorption coefficient, significantly influence the thermomechanical response of the irradiated Si in contrast to the dense metals commonly studied in the literature [25,26].

It is particularly significant to investigate the influence of laser parameters, pulse overlap, material properties, and the boundaries that reflect mechanical waves on LIPSS induction. This is especially critical for Si wafers, given their extensive use in technological devices [32,33]. Building upon this foundation, we numerically investigate the induction of ripples upon ps laser scanning on a Si wafer target exploiting the interpretation of the dynamic thermomechanical behavior of a single ps laser irradiation at predetermined key positions next to reflective boundary planes. The wisely selected focal spot distance from the reflecting boundaries allows for the generation, propagation and reflection of mechanical waves that interfere and act like the modified material fingerprint at a sequential successive laser scanning process.

This work presents a novel thermomechanical study on periodic surface structuring of Si wafers by laser-induced mechanical waves which directly pattern the molten surface, not including the contribution of the electromagnetic interference. Aiming to decode the sequential laser pulse irradiation dynamic effects, ranging from ps to ms timescales, we systematically examine the thermomechanical equivalent of a single-pulse irradiation next to the rectangular solid Si target boundary surfaces. The free boundaries of the solid target effectively model the material modifications ("fingerprints") left by successive laser pulses, providing critical insight into subsequent LIPSS formation. Based on the optimal focal spot separation for LIPSS induction established in [28], we position the laser spot adjacent to a single boundary to simulate simple wave interference. Additionally, we model irradiation equidistant from two adjacent boundaries at a target vertex to generate complex wave patterns. These two thermomechanical configurations—equivalent to sequential irradiation—simulate how material modifications from a leading pulse precondition the surface and govern the mechanical wave dynamics of a subsequent pulse in both time and space. The results establish the controlled excitation of mechanical waves as a fundamental mechanism for LIPSS formation, providing a novel pathway for deterministic laser-based surface patterning.

## 2. Numerical Modeling and Simulation

The solid square Si target FEM model is discretized by a highly refined uniform mesh to accurately capture its dynamic response while and after interacting with Gaussian-profile ps laser pulses and solved in Ls-Dyna [34]. The multiphysics coupled thermal–mechanical FEM models are simulated on the High-Performance Computer (HPC) Advanced Research Information System (ARIS) [35].

*2.1. Mathematical description*

Given the ultrashort nature of the laser pulse, convective and radiative losses are negligible thus laser-silicon interaction is governed by heat conduction. The one-temperature heat conduction equation is:

$$\rho(r,T) C_p(r,T) \frac{\partial T(r,t)}{\partial t} - \nabla[k(r,T)\nabla T(r,t)] = Q(r,t) \quad (1)$$

where *r* is the vector of location, *T is* the temperature, $\rho$ is the mass density, $C_p$, $k$ are the temperature dependent, specific heat at constant pressure and thermal conductivity respectively, while $Q(r,t)$ is the absorbed laser energy per unit volume, per unit time by the sample. The latent heat of melting is also considered when temperature exceeds the melting point. As previously mentioned, the ps pulse has a Gaussian spatiotemporal profile. Consequently, the absorbed volumetric heat flux, $Q(r,t)$, is given by the following equation:

$$Q(r,t) = I_0(t)(1-R) a_b e^{-4ln2(\frac{t^2-t_0^2}{t_0^2})} e^{-(\frac{x^2+y^2}{r_0^2})} e^{-a_b z} \quad (2)$$

where $I_0(t)$ is the temporal profile of the laser intensity, and $r_0$ and $t_0$ are the beam radius (at $1/e^2$ intensity level) and the FWHM laser pulse duration, respectively. The term $(1-R) I_0(t)$ represents the portion of the laser energy that propagates into the target, where $R$ is the surface reflectivity. The coefficient $\alpha_b$ is the absorption coefficient, which depends on temperature, laser wavelength, surface polish, and the doping concentration of the sample. The exponential term $\exp(-\alpha_b z)$ describes the attenuation of the laser intensity with depth $z$.

Rapid local heating and associated thermal expansion generate a stress field, resulting in ultrasonic waves that propagate in the material. The equation of wave propagation is written in the form:

$$\rho(r,T)\frac{\partial^2 U(r,t)}{\partial t^2} = \mu \nabla^2 U(r,T) + (\lambda+\mu)\nabla[\nabla U(r,t)] - a(3\lambda + 2\mu)\nabla T(r,t) \quad (3)$$

where $U$ represents the displacement, $\alpha$ the thermal expansion coefficient and $\lambda$, $\mu$ are Lame constants depending on the material. The mechanical behavior of the Si target can be expressed by the following equations (4, 5) where $\sigma_{ij}$ and $\varepsilon_{ij}$ are the stress and strain tensors in the $ij$ plane respectively and $T_0$ the ambient temperature:

$$\sigma_{ij} = 2\mu\varepsilon_{ij} + \lambda\varepsilon_{kk}\delta_{ij} - (3\lambda + 2\mu)a(T-T_0)\delta_{ij} \quad (4)$$

$$\varepsilon_{ij} = \frac{1}{2}\left(\frac{\partial U_i}{\partial x_j} + \frac{\partial U_j}{\partial x_i}\right) \quad (5)$$

The hydrodynamic and bulk behavior of the solid target is described using the Grüneisen equation of state which gives the pressure as function of volume and internal energy of the material. The equation (6) describes the compressed and the (7) expanding materials:

$$\frac{\rho_0 C^2 \mu_d [1+(1-\frac{\gamma_0}{2})\mu_d - \frac{\alpha_1}{2}\mu_d^2]}{\left[1-(s_1-1)\mu_d - s_2\frac{\mu_d^2}{\mu_d+1} - s_3\frac{\mu_d^3}{(\mu_d+1)^2}\right]^2} + (\gamma_0 + \alpha_1\mu_d)E \quad (6)$$

$$p = \rho_0 C^2 \mu_d + (\gamma_0 + \alpha_1\mu_d)E \quad (7)$$

In these equations $C$ represents the sound speed, $E$ the internal energy per initial volume and $\gamma_0$, $\alpha_1$, $s_1$, $s_2$, $s_3$ are unitless constants. $s_1$, $s_2$, and $s_3$ are the coefficients of the slope in the $u_p$–$u_s$ curve (the shock wave velocity $u_s$ varies linearly, with respect to the particle velocity $u_p$), $\gamma_0$ is the Grüneisen parameter and $\alpha_1$ represents the first order volume correction to $\gamma_0$. The term $\mu_d$ is a volumetric parameter defined by the relation $\mu_d = (\rho/\rho_0) - 1$.

## 2.2. Geometry, discretization & simulation parameters

The 3D FEM model, capable of simulating the response of a Si target under ps laser irradiation, was originally developed and validated for the study of solid Si behavior under ns and ps laser irradiation in [27] and in [28], respectively. Herein, the 3D multiphysics model is modified and adapted to investigate the Si response under single- and sequential- ps laser irradiations, varying in space and in time.

The rectangular workpiece with dimensions of 160 μm × 160 μm × 3 μm (along the X, Y, and Z axes, respectively) is discretized using eight-node hexahedral solid elements of dimensions 1 μm × 1 μm × 75 nm. The simulation region consists of 1024000 elements. This configuration was found to be suitable for describing the thermo-mechanical response of the target across both short (ps) and long (ms) spatiotemporal scales [27,28].

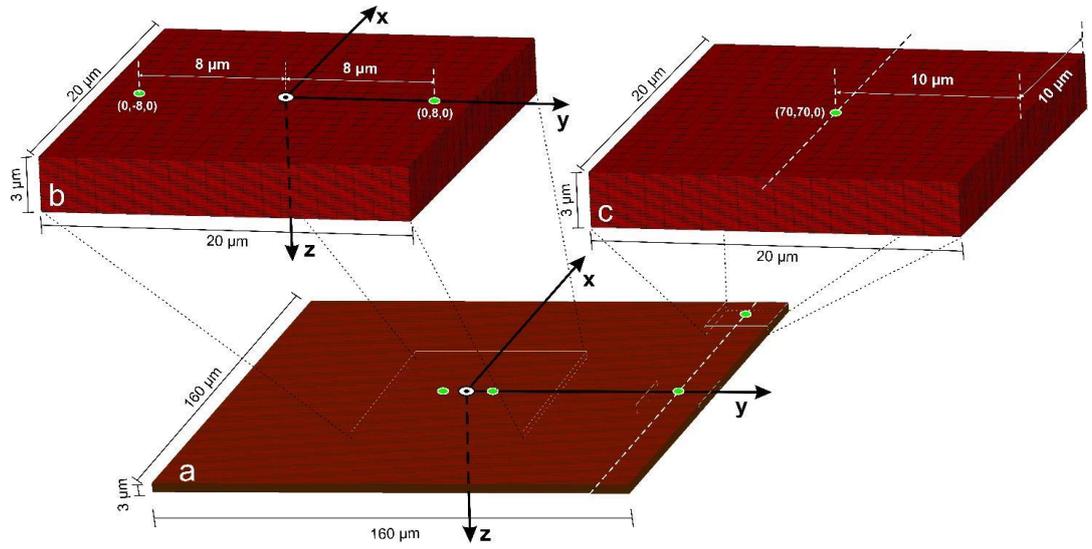

**Figure 2.** Geometry, discretization and modeling parameters of the irradiated Si solid target.

To decipher the complex sequential-pulse phenomena reported in [28], we employ a thermomechanical analog single-pulse model. The study in [28] established that a laser spot center distance below 25 μm is required for LIPSS formation and stress/strain amplification. Accordingly, we simulate a single ps laser pulse focused on point (0,70,0), placing it 10 μm from the right boundary, as shown in Figure 2a. To further investigate the effect of a circular modification boundary from a prior pulse, a second single-pulse simulation is performed, focusing the laser at the upper-right vertex, point (70,70,0), 10 μm from both adjacent edges (see zoomed detail in Figure 2c).

Figure 2a shows the discretized Si model geometry, irradiated on the top surface (xy-plane). The two green spots on either side of the center, positioned 8 μm from the target's origin, mark the centers of two sequential laser pulses with a 25 μm diameter focal spot. The first pulse is focused on point (0,8,0), as shown in the zoomed detail of Figure 2b, while the second pulse is focused at (0,-8,0). The second pulse irradiates the target after a minimum delay of 20 μs, ensuring the target has cooled to ambient temperature. In this sequential-pulse model, the five non-irradiated surfaces have non-reflective boundary conditions, unlike the single-pulse models where the four perimeter boundaries are treated as reflective to simulate the boundaries created by successive pulses.

### 2.3. Material properties and laser parameters

To describe the properties of Si target we used the empirical Johnson–Cook (J-C) [36] material model that provides the flow stress $\sigma$ (Equation 8) and in case of high strain rates includes a fracture model that defines the equivalent plastic strain $\varepsilon_f$ in case of damage (Equations 9,10):

$$\sigma = (A + B\varepsilon^n)\left(1 + C\ln\frac{\dot{\varepsilon}}{\dot{\varepsilon}_0}\right)\left(1 - \left(\frac{T-T_r}{T_m-T_r}\right)^m\right) \quad (8)$$

$$\varepsilon_f = \left(D_1 + D_2 e^{D_3\frac{p}{\sigma_{VM}}}\right)\left(1 + \frac{D_4\ln\dot{\varepsilon}}{\dot{\varepsilon}_0}\right)\left(1 + D_5\frac{T-T_r}{T_m-T_r}\right) \quad (9)$$

$$D = \Sigma\left(\frac{\Delta\varepsilon}{\varepsilon_f}\right) \quad (10)$$

In the equations (8-10) $T$ represents the temperature of the target, $T_r$ the room ambient temperature and $T_m$ the melting point of the material. Parameters $A, B, C, n, m$ are experimental constants which depend on the material, $p$ is the pressure, while $\dot{\varepsilon}$ and $\dot{\varepsilon}_0$ are the strain rate and the reference strain rate, respectively and $D_1$-$D_5$ are the failure parameters of the material. Finally, $D$ is the damage parameter. When $D$ becomes 1, the material fractures. The temperature dependent values of the mechanical properties and thermal properties, as well as the J-C parameters of the Si sample used in the simulations can be found in the supplementary data of [28].

The beam diameter on the sample surface is assumed to be 25 μm (at $1/e^2$ intensity level). In all simulations, the FWHM pulse duration was 6 ps. The reflectivity of Si at 1030 nm is taken for simplicity to be 0.32 at room temperature [37], and the absorption coefficient is temperature dependent [38]. A laser fluence of 1.64 J/cm$^2$ is considered to study the occurring physical phenomena in the melting regime.

The material properties of Si wafers vary depending on dopant concentration, crystal quality, the type of irradiated surface, and rising temperature. The simulations revealed that a critical factor influencing the results was the laser light's penetration depth. In pure Si crystals, the penetration depth can be calculated as $1/\alpha_b$, where $\alpha_b$ is the absorption coefficient for a specific wavelength. However, in Si wafers, the absorption coefficient depends on the proportion of dopants, particularly in the case of highly doped wafers [39-41]. More significantly, it increases substantially with rising temperature. At elevated temperatures, the absorption coefficient can increase dramatically [38,42], making it challenging to accurately calculate the penetration depth, which becomes smaller than the conventional estimate of $1/\alpha_b$ [43]. For a very high dopant Si wafer layered material, the penetration depth can be less than 1 μm for IR laser light [44,45]. For this reason, different mean values of penetration depths were simulated.

## 3. Results

A single-temperature model describes the laser-matter interaction, justified by silicon's characteristic electron-lattice thermalization time of ~10 ps [46,47], which is comparable to the pulse duration. To investigate how the boundary geometry of a leading pulse influences subsequent LIPSS formation, we employ two equivalent single-pulse irradiation models that probe sequential thermomechanical effects: one with the pulse focused near a reflective plane boundary, and another near a reflective corner vertex. Three distinct irradiation scenarios are examined: i) a ps laser pulse applied 70 μm from the target's center and 10 μm from its right edge, ii) a pulse focused 10 μm from both adjacent edges at a corner, iii) two successive, partially overlapping pulses to demonstrate how material modifications from the leading pulse affect LIPSS formation by the second pulse. These simulations are performed for penetration depth (PD) of 75 nm, 150 nm, and 2.5 μm, revealing this parameter's crucial role in LIPSS formation, which depends on material properties and sample characteristics. For single pulses, simulation times were approximately 10 μs (75 nm PD), 20 μs (150 nm PD), and 50 μs (2.5 μm PD) for the target to return to ambient temperature. For sequential pulses, simulation times roughly doubled, except for the 2.5 μm PD case, which required ~500 μs after the second pulse for cooling down.

*3.1. Single-pulse irradiation next to a reflective boundary plane*

The thermal response of the Si target varies significantly with PD. For a PD of 75 nm, the surface temperature peaks at 1800 °C at 20 ps and remains above silicon's melting point (1414 °C) for ~900 ps. The heated region is superficial, with temperatures exceeding 1400 °C in a 22 μm diameter area and higher than 1000 °C confined to a depth of ~150 nm. Consequently, the target cools to ambient temperature within ~9 μs. For a PD of 150 nm, the maximum surface temperature reaches 1430 °C at 20 ps and remains above melting for ~3 ns. A smaller 18 μm diameter region exceeds 1400 °C, but the heat penetrates more deeply, with temperatures higher than 1000 °C reaching ~200 nm. The temperature at a 3 μm depth is 160 °C, compared to 90 °C for the 75 nm PD case. The deeper heating results in a longer cooling time of ~20 μs to ambient temperature. For a PD of 2.5 μm, the surface temperature peaks at 1420 °C at 22 ps, sustaining a molten state for ~100 ns. A 20 μm diameter region exceeds 1400 °C, with temperatures higher than 1300 °C extending to a depth of 2.2 μm. The temperature at 3 μm depth exceeds 1200 °C. As the entire target is heated, it requires ~0.2 ms to reach a uniform thermal equilibrium at 55 °C. The evolution of temperature distribution for PD 75 nm is presented in Figure 3 for three representative temporal moments.

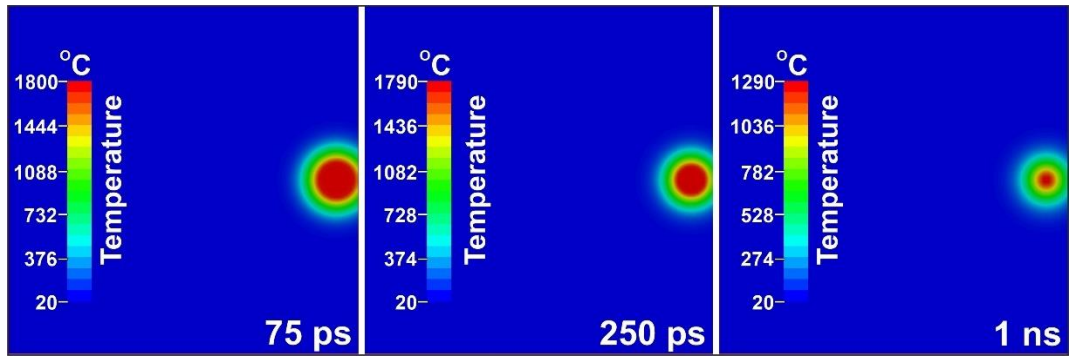

**Figure 3.** Evolution of temperature distribution for PD 75 nm.

Figure 4 presents the evolution of vertical z-displacement following laser irradiation at three different penetration depths. For a PD of 75 nm, interference patterns begin to form at approximately 75 ps, along the y-axis. These are attributed to mechanical waves that propagate and reflect off the free boundary plane. The interference between incoming and reflected waves creates the observed standing waves. These waves establish compression and rarefaction, and the resulting pressure variations displace the molten surface, creating ripples through mechanical wave-driven hydrodynamic flow. By 250 ps, the standing wave pattern occupies a larger area, exhibiting a periodicity close to the laser wavelength (1.03 μm), as observed in literature studies [5,11,16]. The pattern shows fixed nodes and antinodes that oscillate in place while maintaining a consistent wavelength. At 1 ns after irradiation the amplitude of the standing wave patterns has increased. The deformation depth stabilizes within 7.5 μs following irradiation. At 10 μs, when the target has returned to ambient temperature, the maximum deformation depth is 35 nm, within the irradiated area, and distinct ripples remain visible. For the PD value of 150 nm, standing wave patterns emerge by 250 ps and are fully formed across a larger area by 1.3 ns. It is evident that at 2 ns the amplitude of the waves starts to reduce due to the start of dissipation of the mechanical wave energy. At 20 μs after irradiation, the maximum deformation depth is 95 nm, within the irradiated area. Simultaneously, faint ripples persist after solidification. For PD 2.5 μm, higher amplitude standing waves form ripples by 2.5 ns, expanding by 6.5 ns. However, as can be seen at 10 ns these incipient features are weakened by hydrodynamic flow within the extensive molten pool, which persisted for ~100 ns. Moreover at 10 ns an ultrasonic wave is also evident and in accordance with the detailed study published in [27]. The maximum deformation depth is 450 nm, found within the irradiated area and achieved when the Si target reaches a thermodynamic equilibrium at 100 μs.

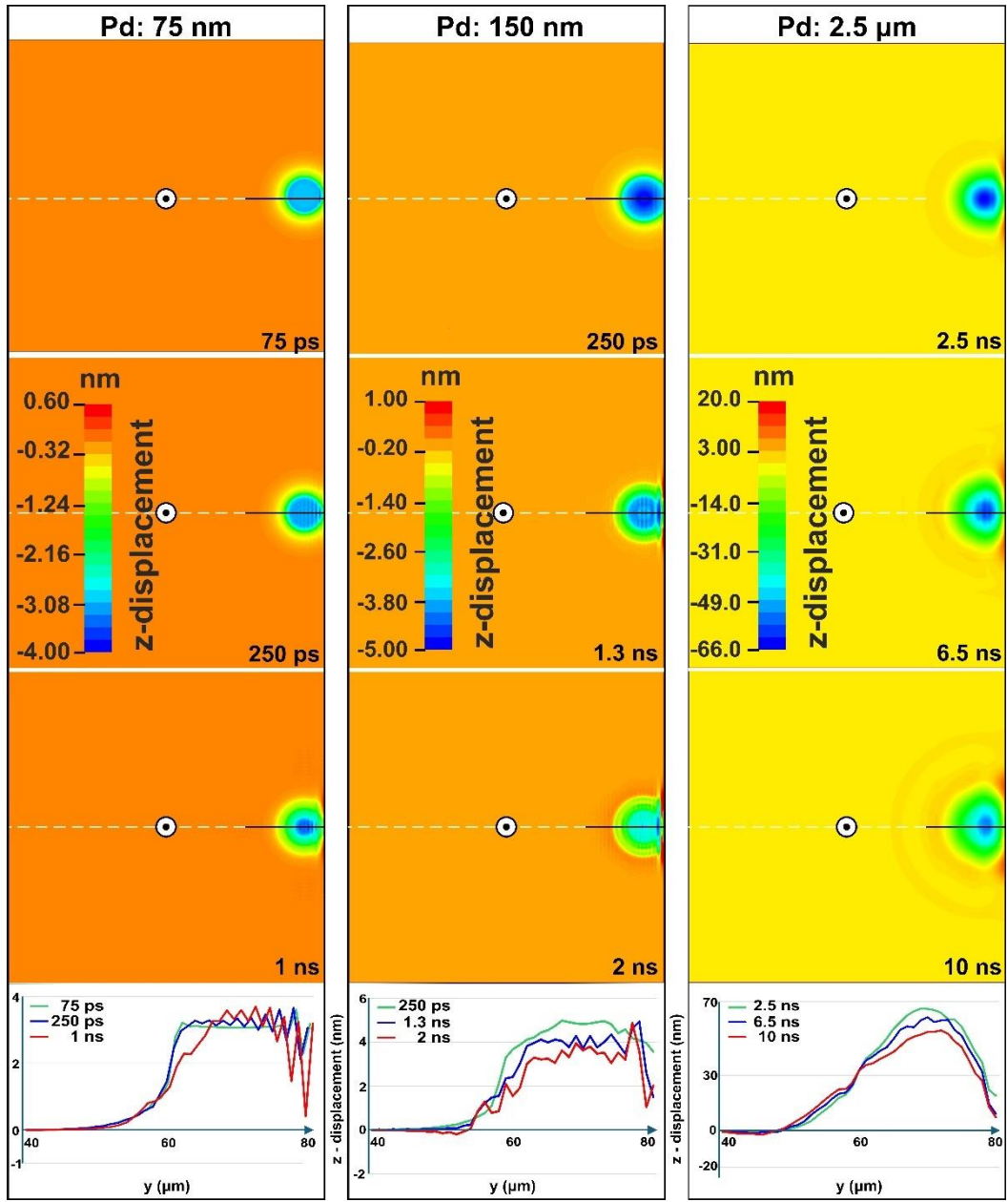

**Figure 4.** Contour plots for the evolution of the vertical z-displacement for three different PDs and the graphs of the corresponding lineouts.

For a PD of 75 nm, the Von Mises stress peaked at approximately 1.7 GPa at 25 ps, dropped to 0.8 GPa by 500 ps, and subsequently increased to a final value of 1.6 GPa (end of simulation time). For a PD of 150 nm, the stress fluctuated between 1.1 GPa and 1.8 GPa before stabilizing at a final value of 1.6 GPa by 10 µs. For a PD of 2.5 µm, the stress varied between 1.1 GPa and 1.8 GPa over a longer duration (~50 µs), ultimately stabilizing at a lower residual stress of 1 GPa. Figure 5 demonstrates the spatial distribution of the residual stresses and strains, in the irradiation region, at the end of simulation time (10 µs for 75 nm PD, 20 µs for 150 nm PD and 50 µs for 2.5 µm PD. The von Mises stress results indicate that for lower PD (75 nm and 150 nm), high residual stress is intensely concentrated at the boundary of the laser spot. In contrast, the deeper penetration depth (2.5 µm) creates a broader zone of material modification with a lower peak stress. For a PD of 75 nm, the residual plastic strain within a central 10 µm diameter area is approximately 0.027–0.033. For a PD of 150 nm, the strain is significantly higher (0.06–0.076) but confined to a smaller central region of 5 µm diameter. For a PD of 2.5 µm, a substantial plastic strain (0.02–0.04) is sustained across the entire irradiated zone.

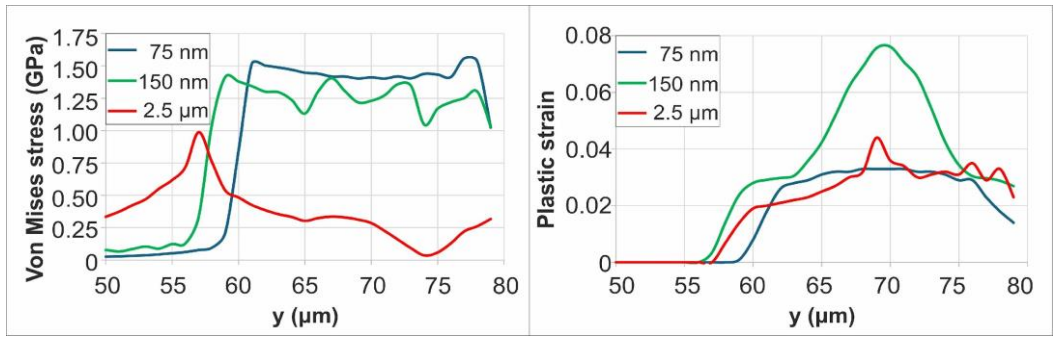

**Figure 5.** Residual stress and plastic strain distributions for varying PD.

*3.2. Single-pulse irradiation next to a reflective boundary vertex*

Irradiation close to a samples boundary vertex consistently yielded slightly lower maximum temperature, von Mises stress, and residual plastic strain compared to the straight-edge case, regardless of the PD. Figure 6 presents the evolution of vertical displacement for irradiation near a corner at three different PDs. For a PD of 75 nm, interference patterns emerge by 80 ps, demonstrating the initial formation of mechanical standing waves. Well-defined periodic structures are established by 250 ps, with the amplitude peaking around 1 ns before beginning to decay after 1.5 ns due to energy dissipation. For a PD of 150 nm, wave patterns emerge by 100 ps. The amplitude increases significantly by 700 ps compared to the 150 ps state but is substantially reduced by 1.5 ns. For a PD of 2.5 µm, higher-amplitude standing waves form ripples by 2.2 ns but are subsequently weakened by hydrodynamic flow in the extensive molten pool—behavior consistent with the straight-edge case. Although acoustic interference patterns re-emerged transiently during solidification, they ultimately failed to stabilize into permanent surface structures.

Critically, vertical displacements were consistently smaller during corner irradiation. The doubly reflected waves interfere in a manner that reduces out-of-plane motion while minimally affecting longitudinal displacements. Although the final displacement magnitudes were lower (30 nm, 40 nm, and 230 nm for penetration depths of 75 nm, 150 nm, and 2.5 µm, respectively), the transient fluctuations were more intense. This demonstrates that while the overall deformation is reduced, the dynamic process of ripple formation is amplified by the complex wave interference from both edges.

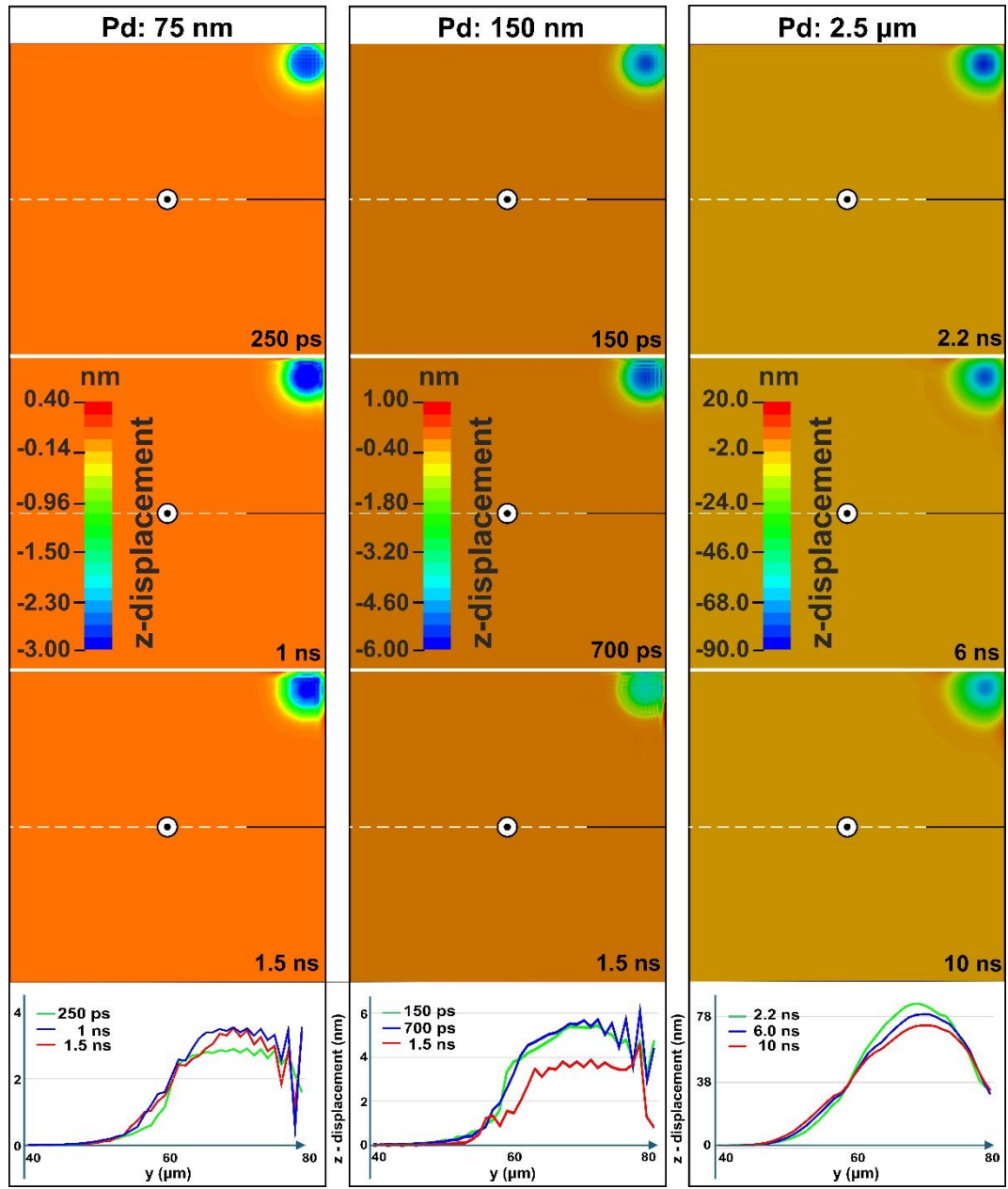

**Figure 6.** Contour plots for the evolution of the vertical z-displacement for three different PDs and the graphs of the corresponding lineouts.

Figure 7 demonstrates the spatial distribution of residual stresses and strains within the irradiation region near the corner, at the end of simulation time (10 μs for 75 nm PD, 20 μs for 150nm PD and 50 μs for 2.5 μm PD). Consistent with the earlier observation, this configuration exhibited slightly lower values than the straight-edge case. The von Mises stress results align with the findings from Section 3.1: for shallow penetration depths (75 nm and 150 nm), high residual stress is intensely concentrated at the boundary of the laser spot. In contrast, the deeper penetration depth (2.5 μm) creates a broader zone of material modification with a lower peak stress. In contrast, the deeper PD (2.5 μm) creates a broader zone of material modification with a lower peak stress. For a PD of 75 nm, the residual plastic strain within a central 10 μm diameter area is approximately 0.023–0.03. For a PD of 150 nm, the strain is significantly higher (0.055–0.07) but confined to a smaller central region of 5 μm diameter. For a PD of 2.5 μm, lower plastic strains (0.015–0.018) are observed in the irradiated area.

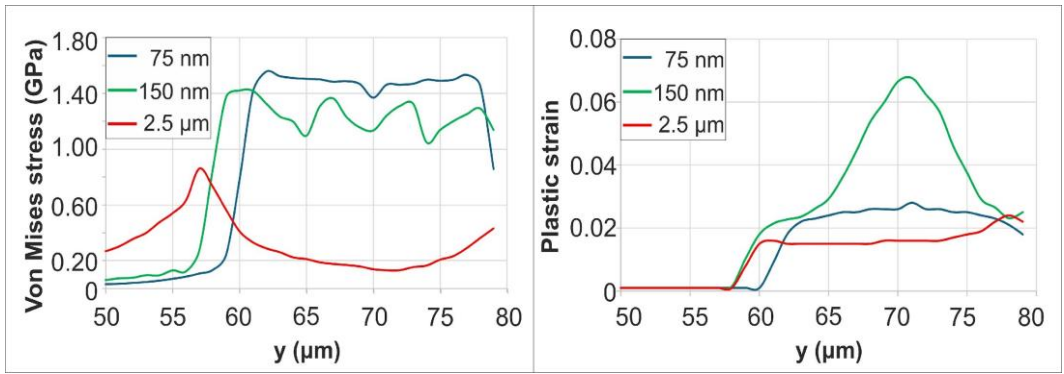

**Figure 7.** Residual stress and plastic strain distributions for varying PD.

Based on the results of sections 3.1 and 3.2, we conclude that a smaller PD yields a higher maximum temperature, a shorter cooling time, and reduced vertical displacements. While ripple formation is initiated for all cases, a deep molten pool at the 2.5 µm PD delays their clear formation. Subsequently, these ripples are depleted by extensive hydrodynamic flow as the material reorganizes.

*3.3. Sequential laser pulses irradiation*

To investigate the influence of the material modifications induced from the leading pulses on the following, we simulated the irradiation sequence of two pulses, by focusing the second laser pulse at the distances of 24, 20, 16 and 12 µm on the y-axis, from the first focal spot center, after targets' cool down to the ambient temperature. The results established a separation threshold: for distances greater than 20 µm, the first pulse had a negligible effect. Conversely, a 12 µm separation resulted in excessive overlap, obscuring individual pulse effects. Consequently, the 16 µm separation distance is further analyzed as a representative case that exhibits clear thermomechanical interaction without significant spatial overlap. These findings agree with the experimental results presented in [28] and are in accordance with the distance of 10 µm applied between the single-pulse irradiation focal spot center from targets boundaries.

For a PD of 75 nm, the first pulse induced a peak temperature of 1805 °C at 18 ps, which was sustained until 100 ps. The shallow penetration depth confined most thermal energy to a depth of approximately 150 nm. The material within a central 10 µm diameter region remained above the melting point for ~1 ns. The thermomechanical response produced a maximum von Mises stress of approximately 1.6 GPa at 25 ps, and a plastic strain of 0.033 on the top surface. A second pulse was applied 10 µs later, after the target cooled to ambient temperature, with its focal point offset by 16 µm. This pulse produced nearly identical thermal behavior, reaching a similar peak temperature of 1800 °C. As in the single-pulse case, the target cooled to ambient temperature within ~10 µs.

The mechanical response, however, was significantly more pronounced. The second irradiated area experienced higher stress, with maximum plastic strain and von Mises stress reaching 0.037 and 1.62 GPa, respectively, exceeding the first pulse's values. Figure 8 shows the evolution of von Mises stress and plastic strain distribution following the first and second pulses. Once the target has cooled to ambient temperature (10 µs after the second pulse), the effect of the successive, partially overlapping pulses remains more pronounced, as evidenced by the higher residual stress and strain fields. Furthermore, the maximum deformation depth was ~38 nm after the first pulse and increased to 55 nm following the second pulse.

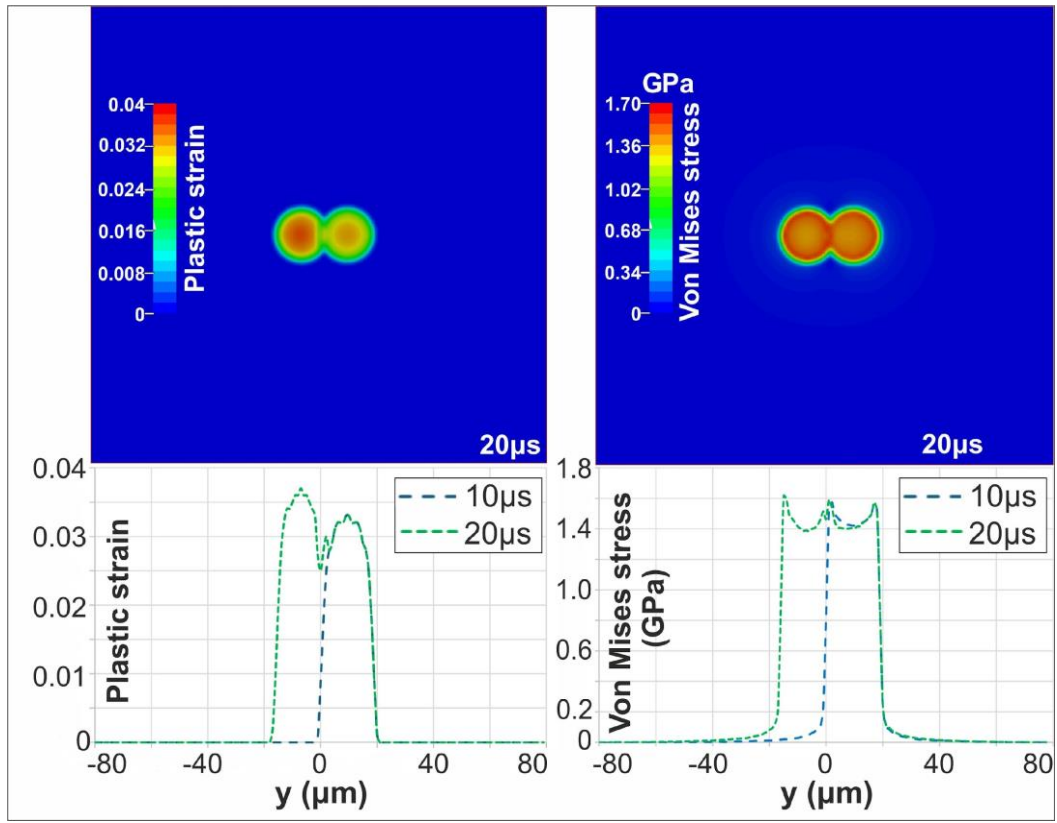

**Figure 8.** Contour plots of plastic strain and Von Mises stress at 20 μs (top); Evolution of plastic strain and Von Mises stress following the first and second pulses (bottom).

For a PD of 150 nm, the first pulse produced a peak temperature of 1430 °C at 18 ps. The second pulse resulted in nearly identical thermal behavior, reaching the same maximum temperature and heated depth. The mechanical response, however, was significantly more pronounced, indicating an increased sensitivity to the material and surface properties modified by the first pulse. During the second pulse, the maximum von Mises stress and plastic strain increased to approximately 1.54 GPa and 0.068, respectively, exceeding the first-pulse values of 1.48 GPa and 0.06. As in the single-pulse case, the target cooled to ambient temperature within approximately 20 μs. Figure 9 shows the evolution of the Von Mises stress and plastic strain distribution following the first and second pulses, once the target has cooled to ambient temperature (20 μs after the second pulse). Furthermore, the maximum deformation depth was ~90 nm after the first pulse and increased to 110 nm following the second pulse.

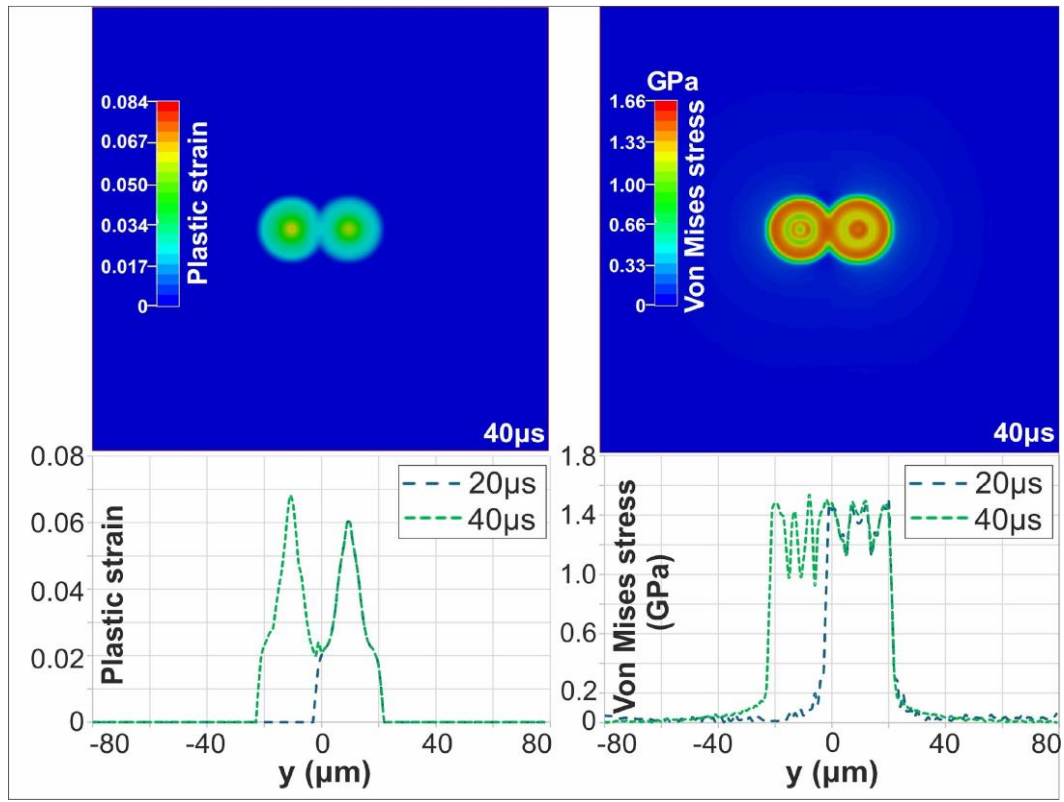

**Figure 9.** Contour plots of plastic strain and Von Mises stress at 40 µs (top); Evolution of plastic strain and Von Mises stress following the first and second pulses (bottom).

For a PD of 2.5 µm, the first pulse produced thermomechanical behavior consistent with single-pulse irradiation, reaching a similar maximum temperature and heated depth. The second pulse, applied after a 50 µs delay, generated a more pronounced mechanical response due to the localized material modifications from the first pulse. Figure 10 depicts the plastic strain distribution following the first and second pulses. A midplane sectional view reveals the strain profile beneath the laser spots. In contrast to all previous cases, the maximum residual plastic strain of 0.041 is located on the subsurface, not on the top surface. Furthermore, the surface plastic strain increased to 0.018, exceeding the first-pulse value of 0.012. The target required 550 µs to cool to ambient temperature following the second pulse.

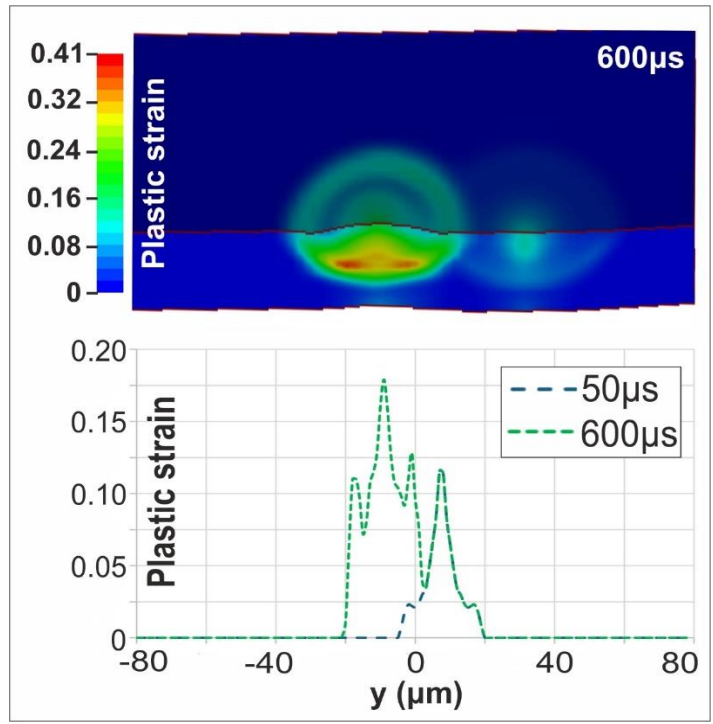

**Figure 10.** Cross-sectional plastic strain distribution along the symmetry plane at 600 μs (top) Evolution of the surface plastic strain following the first and second pulses along the y-axis (bottom).

Figure 11 presents characteristic vertical displacement profiles for lower (150 nm) and higher (2.5 μm) PD at 2 ns and 10 ns, respectively, after the second pulse. At these times, ripple formation is evident within the molten region. The ripple formation closely resembles the morphology and mechanism seen in single-pulse cases (Figures 3 and 5), confirming that the same fundamental process of mechanical standing wave interference governs the initial stage of ripple formation. A distinct bulge is observed between -20 μm and 0 μm for the higher PD case, indicating that the thermal and mechanical influence of the second pulse persists at 10 ns. This contrasts with the lower PD case, where the material has largely solidified, demonstrating how PD governs the temporal evolution of surface dynamics. The maximum deformation depth at the end of the simulation reaches ~600 nm.

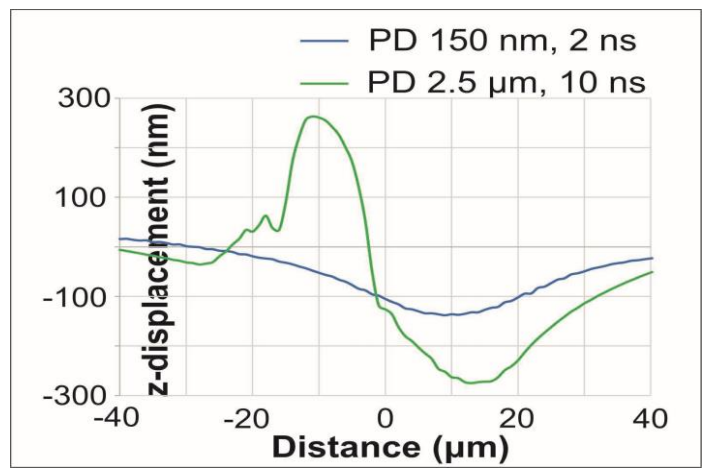

**Figure 11.** Vertical displacement profiles for lower and higher PD at 2 ns and 10 ns, respectively, after the second pulse.

The comparison between the single- and sequential- pulse irradiation cases reveals a clear amplification of thermomechanical effects. In the single-pulse simulations, ripple formation arises from the interference of thermoelastic waves reflecting from nearby boundaries, leading to well-defined standing wave patterns for shallow PDs (75–150 nm) and partial suppression at deeper PDs

(~2.5 µm) due to prolonged melting and hydrodynamic flow. When a second pulse is applied after full thermal relaxation, as in the two-pulse configuration, the pre-existing residual stress and strain fields established by the first pulse significantly enhance the subsequent deformation response. The second pulse produces higher von Mises stresses, larger plastic strain amplitudes, and deeper surface modulation compared to the single-pulse case. This amplification effect confirms that the single-pulse analogues near reflective boundaries successfully reproduce the mechanical conditions of successive irradiation, while the sequential pulse simulations validate laser scanning strengthens standing wave interference and promotes the persistence and regularity of LIPSS features.

## 4. Discussion and Conclusions

This work provides a comprehensive thermomechanical interpretation of the mechanisms governing laser-induced periodic surface structure formation on Si wafers irradiated by ps laser pulses. Through detailed finite element multiphysics simulations, the dynamic interaction between thermal, mechanical, and hydrodynamic phenomena was examined under varying PDs, boundary geometries, and irradiation sequences. The results confirm that mechanical standing wave interference in the molten phase plays a dominant role in the initial formation and stabilization of surface ripples.

The simulations revealed that PD is a critical parameter determining both the spatial and temporal evolution of surface morphology. Shallow PDs (75–150 nm) generate strong, localized thermal gradients and short-lived molten layers that sustain standing mechanical waves capable of imprinting quasi-periodic ripples with periodicities close to the laser wavelength. In contrast, deeper PDs (~2.5 µm) yield extended molten regions with long cooling times, where hydrodynamic motion dominates and smooths out transient surface structures. This highlights PD as a key parameter for tailoring LIPSS morphology via laser wavelength, fluence, or wafer doping.

The influence of boundary geometry was also evident. When irradiation occurred near a single edge, reflected acoustic waves interfered constructively with incoming stress waves, generating well-defined standing wave patterns and surface ripples of significant amplitude. Near a corner, the superposition of doubly reflected waves reduced overall out-of-plane deformation but amplified transient stress oscillations, implying that the spatial confinement of the mechanical field can be exploited to modulate the ripple amplitude and uniformity.

Sequential double-pulse simulations demonstrated the cumulative nature of thermomechanical effects. When the second pulse was applied after full thermal relaxation, residual stresses and plastic strains from the first pulse acted as pre-conditioning factors that enhanced local energy absorption and wave amplitude during the subsequent irradiation. This cumulative amplification of stress and strain led to larger deformation depths compared to single-pulse cases. Coherent interference patterns that resemble experimentally observed LIPSS [28] were computed.

These results shed light on the interplay between residual mechanical fields and re-solidification dynamics. Future work may extend this framework by integrating electromagnetic field coupling and two-temperature models to capture the early-stage carrier dynamics, validated by experimental measurements, to further advance the deterministic control of laser-induced nanostructuring processes for Si-based photonic and microelectronic applications.

## References

1. Bonse, J.; Krüger, J.; Höhm, S.; Rosenfeld, A. Femtosecond Laser-Induced Periodic Surface Structures. *J. Laser Appl.* **2012**, *24*, 042006.
2. Almeida, G.F.B.; Martins, R.J.; Otuka, A.J.G.; Siqueira, J.P.; Mendonca, C.R. Laser Induced Periodic Surface Structuring on Si by Temporal Shaped Femtosecond Pulses. *Opt. Express* **2015**, *23*, 27597-27607.
3. Zehetner, J.; Schmidmayr, D.; Piredda, G.; Kasemann, S.; Matylitskaya, V.; Lucki, M. Laser Generated Micro-and-Nanostructures and the Transfer to Polymers for Experimental Use. In Proceedings of the APCOM 2015, Strbske Pleso, Slovakia, 24-26 June 2015.
4. Navickas, M.; Grigutis, R.; Jukna, V.; Tamošauskas, G.; Dubietis, A. Low Spatial Frequency Laser-Induced Periodic Surface Structures in Fused Silica Inscribed by Widely Tunable Femtosecond Laser Pulses. *Sci. Rep.* **2022**, *12*, 20231.

5. Bonse, J. Quo Vadis LIPSS?—Recent and Future Trends on Laser-Induced Periodic Surface Structures. *Nanomaterials* **2020**, *10*, 1950.
6. Bonse, J.; Höhm, S.; Kirner, S.V.; Rosenfeld, A.; Krüger, J. Laser-Induced Periodic Surface Structures—A Scientific Evergreen. *IEEE J. Sel. Top. Quantum Electron.* **2017**, *23*, 9000615.
7. Young, J.F.; Preston, J.S.; van Driel, H.M.; Sipe, J.E. Laser-Induced Periodic Surface Structure. II. Experiments on Ge, Si, Al, and Brass. *Phys. Rev. B* **1983**, *27*, 1155-1172.
8. Dostovalov, A.V.; Korolkov, V.P.; Babin, S.A. Formation of Thermochemical Laser-Induced Periodic Surface Structures on Ti Films by a Femtosecond IR Gaussian Beam: Regimes, Limiting Factors, and Optical Properties. *Appl. Phys. B* **2017**, *123*, 30.
9. Vorobyev, A.Y.; Guo, C. Colorizing Metals with Femtosecond Laser Pulses. *Appl. Phys. Lett.* **2008**, *92*, 041914.
10. Mastellone, M.; et al. LIPSS Applied to Wide Bandgap Semiconductors and Dielectrics: Assessment and Future Perspectives. *Materials* **2022**, *15*, 1378.
11. Bonse, J.; Gräf, S. Maxwell Meets Marangoni—A Review of Theories on Laser-Induced Periodic Surface Structures. *Laser Photonics Rev.* **2020**, *14*, 2000215.
12. Ince, F.D.; Ozel, T. Laser Surface Texturing of Materials for Surface Functionalization: A Holistic Review. *Surf. Coat. Technol.* **2025**, *498*, 131818.
13. Fuentes-Edfuf, Y.; Sanchez-Gil, J.A.; Florian, C.; Giannini, V.; Solis, J.; Siegel, J. Surface Plasmon Polaritons on Rough Metal Surfaces: Role in the Formation of Laser-Induced Periodic Surface Structures. *ACS Omega* **2019**, *4*, 6939-6946.
14. Tsibidis, G.D.; Fotakis, C.; Stratakis, E. From Ripples to Spikes: A Hydrodynamical Mechanism to Interpret Femtosecond Laser-Induced Self-Assembled Structures. *Phys. Rev. B* **2015**, *92*, 041405.
15. Gurevich, E.L. Mechanisms of Femtosecond LIPSS Formation Induced by Periodic Surface Temperature Modulation. *Appl. Surf. Sci.* **2016**, *374*, 56-60.
16. Höhm, S.; Herzlieb, M.; Rosenfeld, A.; Krüger, J.; Bonse, J. Dynamics of the Formation of Laser-Induced Periodic Surface Structures (LIPSS) upon Femtosecond Two-Color Double-Pulse Irradiation of Metals, Semiconductors, and Dielectrics. *Appl. Surf. Sci.* **2016**, *374*, 331-338.
17. Bonse, J.; Krüger, J. Pulse Number Dependence of Laser-Induced Periodic Surface Structures for Femtosecond Laser Irradiation of Silicon. *J. Appl. Phys.* **2010**, *108*, 034903.
18. Nivas, J.J.; Anoop, K.K.; Bruzzese, R.; Philip, R.; Amoruso, S. Direct Femtosecond Laser Surface Structuring of Crystalline Silicon at 400 nm. *Appl. Phys. Lett.* **2018**, *112*, 121601.
19. Ali, A.; Piatkowski, P.; Alnaser, A.S. Study on the Origin and Evolution of Femtosecond Laser-Induced Surface Structures: LIPSS, Quasi-Periodic Grooves, and Aperiodic Micro-Ridges. *Materials* **2023**, *16*, 2184.
20. Gao, Y.-F.; Yu, C.-Y.; Han, B.; Ehrhardt, M.; Lorenz, P.; Xu, L.-F.; Zhu, R.-H. Picosecond Laser-Induced Periodic Surface Structures (LIPSS) on Crystalline Silicon. *Surf. Interfaces* **2020**, *19*, 100538.
21. Trtica, M.S.; Gakovic, B.M.; Radak, B.B.; Batani, D.; Desai, T.; Bussoli, M. Periodic Surface Structures on Crystalline Silicon Created by 532 nm Picosecond Nd:YAG Laser Pulses. *Appl. Surf. Sci.* **2007**, *254*, 1377-1381.
22. Grigoryeva, M.S.; Kutlubulatova, I.A.; Lukashenko, S.Y.; Fronya, A.A.; Ivanov, D.S.; Kanavin, A.P.; Timoshenko, V.Y.; Zavestovskaya, I.N. Modeling of Short-Pulse Laser Interactions with Monolithic and Porous Silicon Targets with an Atomistic–Continuum Approach. *Nanomaterials* **2023**, *13*, 2809.
23. Yang, J.; Zhang, D.; Wei, J.; Shui, L.; Pan, X.; Lin, G.; Sun, T.; Tang, Y. The Effect of Different Pulse Widths on Lattice Temperature Variation of Silicon under the Action of a Picosecond Laser. *Micromachines* **2022**, *13*, 1119.
24. Kan, Z.; Zhu, Q.; Ren, H.; Shen, M. Femtosecond Laser-Induced Thermal Transport in Silicon with Liquid Cooling Bath. *Materials* **2019**, *12*, 2043.
25. Dimitriou, V.; Kaselouris, E.; Orphanos, Y.; Bakarezos, M.; Vainos, N.; Tatarakis, M.; Papadogiannis, N.A. Three-Dimensional Transient Behavior of Thin Films Surface under Pulsed Laser Excitation. *Appl. Phys. Lett.* **2013**, *103*, 114104.
26. Kaselouris, E.; Nikolos, I.K.; Orphanos, Y.; Bakarezos, E.; Papadogiannis, N.A.; Tatarakis, M.; Dimitriou, V. Elastoplastic Study of Nanosecond-Pulsed Laser Interaction with Metallic Films Using 3D Multiphysics FEM Modeling. *Int. J. Damage Mech.* **2016**, *25*, 42–55
27. Papadaki, H.; Kaselouris, E.; Bakarezos, M.; Tatarakis, M.; Papadogiannis, N.A.; Dimitriou, V. A Computational Study of Solid Si Target Dynamics under ns Pulsed Laser Irradiation from Elastic to Melting Regime. *Computation* **2023**, *11*, 240.

28. Mirza, I.; Sládek, J.; Levy, Y.; Bulgakov, A.V.; Dimitriou, V.; Papadaki, H.; Kaselouris, E.; Gečys, P.; Račiukaitis, G.; Bulgakova, N.M. Coherence Effects in LIPSS Formation on Silicon Wafers upon Picosecond Laser Pulse Irradiations. *J. Phys. D: Appl. Phys.* **2025**, *58*, 085307.
29. Duffy, W.; Bassan, M. Acoustic quality factor of 95%-density molybdenum and its application to high frequency cryogenic gravity-wave antennas, *Cryogenics* **1998**, *38*, 757-762.
30. Duffy, W. Acoustic quality factor of titanium from 50 mK to 300 K, *Cryogenics* **2000**, *40*, 417-420.
31. Buser, R.A.; De Rooij, N.F. Very high Q-factor resonators in monocrystalline silicon, *Sens. Actuator A-Phys.* **1990**, *21*, 323-327.
32. Priolo, F.; Gregorkiewicz, T.; Galli, M.; Krauss, T.F. Silicon Nanostructures for Photonics and Photovoltaics. *Nat. Nanotechnol.* **2014**, *9*, 19–32.
33. Vaghasiya, H.; Miclea, P.T. Investigating Laser-Induced Periodic Surface Structures (LIPSS) Formation in Silicon and Their Impact on Surface-Enhanced Raman Spectroscopy (SERS) Applications. *Optics* **2023**, *4*, 538–550.
34. Hallquist, J. LS-DYNA Theory Manual; Livermore Software Technology Corporation: Livermore, CA, USA, 2006.
35. Aris Documentation. Available online: http://doc.aris.grnet.gr/system/hardware/ (accessed on 09 November 2025).
36. Orphanos, Y.; Kosma, K.; Kaselouris, E.; Vainos, N.; Dimitriou, V.; Bakarezos, M.; Tatarakis, M.; Papadogiannis, N.A. Integrated nanosecond laser full-field imaging for femtosecond laser-generated surface acoustic waves in metal film-glass substrate multilayer materials. *Appl. Phys. A* **2019**, *125*, 269.
37. Schinke, C.; Peest, P.C.; Schmidt, J.; Brendel, R.; Bothe, K.; Vogt, M.R.; Kröger, I.; Winter, S.; Schirmacher, A.; Lim, S.; Nguyen, H.T.; MacDonald, D. Uncertainty analysis for the coefficient of band-to-band absorption of crystalline silicon. *AIP Adv.* **2015**, *5*, 067168.
38. Sin, E.H.; Ong, C.K.; Tan, H.S. Temperature dependence of interband optical absorption of silicon at 1152, 1064, 750, and 694 nm. *Phys. Status Solidi A* **1984**, *85*, 199-204.
39. Jellison, G.E., Jr.; Modine, F.A.; White, C.W.; Wood, R.F.; Young, R.T. Optical Properties of Heavily Doped Silicon between 1.5 and 4.1 eV. *Phys. Rev. Lett.* **1981**, *46*, 1414-1417.
40. Baker-Finch, S.C.; McIntosh, K.R.; Yan, D.; Fong, K.C.; Kho, T.C. Near-infrared free carrier absorption in heavily doped silicon. *J. Appl. Phys.* **2014**, *116*, 063106.
41. Jain, S.C.; Nathan, A.; Briglio, D.R.; Roulston, D.J.; Selvakumar, C.R.; Yang, T. Band-to-band and free-carrier absorption coefficients in heavily doped silicon at 4 K and at room temperature. *J. Appl. Phys.* **1991**, 69, 3687-3690.
42. Kovalev, D.; Polisski, G.; Ben-Chorin, M.; Diener, J.; Koch, F. The temperature dependence of the absorption coefficient of porous silicon. *J. Appl. Phys.* **1996**, *80*, 5978-5983.
43. Hecht, E. Optics, 5th ed.; Pearson Education: London, UK, 2016; p. 137.
44. Tong, Z.; Bu, M.; Zhang, Y.; Yang, D.; Pi, X. Hyperdoped silicon: Processing, properties, and devices. *J. Semicond.* **2022**, *43*, 093101.
45. Kovalev, M.; Nastulyavichus, A.; Podlesnykh, I.; Stsepuro, N.; Pryakhina, V.; Greshnyakov, E.; Serdobintsev, A.; Gritsenko, I.; Khmelnitskii, R.; Kudryashov, S. Au-Hyperdoped Si Nanolayer: Laser Processing Techniques and Corresponding Material Properties. *Materials* **2023**, *16*, 4439.
46. Yang, J.; Zhang, D.; Wei, J.; Shui, L.; Pan, X.; Lin, G.; Sun, T.; Tang, Y. Effect of Different Pulse Widths on Lattice Temperature Variation of Silicon under Picosecond Laser. *Micromachines* **2022**, *13*(7), 1119.
47. Wörle, M.; Holleitner, A.W.; Kienberger, R.; Iglev, H. Ultrafast Hot-Carrier Relaxation in Silicon Monitored by Phase-Resolved Transient Absorption Spectroscopy. *Phys. Rev. B* **2021**, *104*, L041201.